# Dynamic Computation of Runge-Kutta's Fourth-Order Algorithm for First and Second Order Ordinary Differential Equation Using Java

Adesola O. Anidu, Samson A. Arekete, Ayomide O. Adedayo and Adekunle O. Adekoya

Department of Computer Science, Redeemer's University, Ede, Nigeria

**Abstract**

Differential equations arise in mathematics, physics, medicine, pharmacology, communications, image processing and animation, etc. An Ordinary Differential Equation (ODE) is a differential equation if it involves derivatives with respect to only one independent variable which can be studied from different perspectives; such as: analytical methods, graphical methods and numerical methods. This research paper therefore revises the standard Runge-Kutta fourth order algorithm by using compiler techniques to dynamically evaluate the inputs and implement the algorithm for both first and second order derivatives of the ODE. We have been able to develop and implement the software that can be used to evaluate inputs and compute solutions (approximately and analytically) for the ODE function at a more efficient rate than the traditional method.

## 1. Introduction

Computational problems are problems found in areas of study like mathematics, computer and related subjects which can be represented in forms like algorithms, flowcharts and so on. There are various algorithms for solving a wide range of problems; and one of those problems that an algorithm can be represented in is the class of Runge-Kutta methods. Runge-Kutta (RK) methods are a class of methods mostly used for solving IVPs (Initial Value Problems) numerically, because of their speed and accuracy. It is an essential family of implicit and explicit iterative methods needed for the approximation of solutions of ordinary differential equations which was developed around 1900 by German Mathematicians C.Runge and M.W. Kutta. Despite the fact that RK methods require less computation of higher order derivatives, they still give immense accuracy.

However, the set of explicit RK methods for numerically solving IVPs are the most popular because of their speed and accuracy; in which the simplest and most basic method for solving IVPs is the Euler's method also known as Forward Euler [6]. Forward Euler is very easy to understand and implement but it is not as efficient as some higher-order explicit Runge-Kutta methods. These higher-order explicit RK methods operate in a similar fashion to Forward Euler in that they approximate the solution to $y(t)$ by stepping to $t_f$. The difference is that in each step, instead of using just $f(t_n, y_n)$, higher-order explicit Runge-Kutta methods take a weighted average of several function evaluations, typically within $[t_n, t_{n+1}]$. Although this means more computations per step, the accuracy of the solution is much better, relative to the amount of work done. The more efficient explicit Runge-Kutta methods are also a bit harder to understand and implement than Forward Euler but the overall gain in efficiency makes implementing them worth this effort.

Among the class of RK algorithm is the fourth order method which is the most popular often referred to as "RK4" and is also used for solving Initial Value Problems (IVPs). The fourth order Runge-Kutta method is the most powerful of the entire explicit RK methods, so accurate that most computer packages such as Java and C++ use it to compute solutions, numerically for differential equations. An IVP is composed of an Ordinary Differential Equation (ODE) and a prescribed initial value at a certain time. Certain conditions under which an IVP has a unique solution can be known but obtaining this solution in an analytical form can be too cumbersome. However, there are ways to approximate the solutions of most IVPs which is by solving them numerically; obtaining the numerical solution to a problem involves the use of a numerical method. Numerical methods can often produce a solution to any degree of accuracy that the computer can represent.

Arising from the existing limitations of implementing Runge-Kutta fourth order method, this method of solving RK4 poses a problem because the user can only perform operations on the already specified derivative of y or keeps changing the derivative of y for every RK4 problem. This limits the user of the program and is not flexible. The RK4 method for numerically solving ODE's is represented mathematically as:

$$y' = f(x, y).$$

This research work aims to dynamically implement and evaluate the Runge-Kutta fourth order algorithm using compiler techniques for a number of users using Java programming language.

The rest of the paper is organized as follows. Section 2 presents the review of related literature. In section 3 we present the system design, section 4 presents the implementation techniques while section 5 presents the results obtained.

## 2. Literature Review

Runge-Kutta formula is among the oldest and best understood schemes in numerical analysis. Owing to the evolution of a vast and comprehensive body of knowledge, Runge-Kutta still continues to be a source of active research [2]. The most suitable way of solving most initial value problems for a system of ordinary differential equations are mostly provided by Runge-Kutta methods sometimes referred to as "RK" methods. This is based on some reasons; Firstly, Runge-Kutta methods are convergent given that the approximate solution approaches the exact solution. Secondly, they are accurate due to the closeness between the approximate solution and the exact solution. Finally, they are conditionally stable in that it is stable for some values of the parameter.

The first order Runge-Kutta method which is in point of fact the Euler's method is given as:

$$y_{n+1} = y_n + hf(x_n, y_n) + O(h^2) \quad (1)$$

First order RK[7]

This order advances a solution from $x_n$ to $x_{n+1} \equiv x_n + h$ that is, through an interval $h$ which is unsymmetrical. Therefore this method is neither accurate nor stable.

The second order Runge-Kutta method which is in fact the Modified Euler's method is given as:

$$k_1 = hf(x_n, y_n)$$
$$k_2 = hf(x_n + h, y_n + k_1)$$
$$y_{n+1} = y_n + \frac{1}{2}(k_1 + k_2) + O(h^3) \quad (2)$$

Second order RK[7]

Eq. 2 above is also called the midpoint method and as denoted by the error term rules out the first order error term by symmetrization, and in that way, the method becomes second order.

The third order Runge-Kutta method which is the Runge's method is given as:

$$k_1 = hf(x_n, y_n)$$
$$k_2 = hf\left(x_n + \frac{h}{2}, y_n + \frac{k_1}{2}\right)$$
$$k_3 = hf(x_n + h, y_n + k_1)$$
$$y_{n+1} = y_n + \frac{1}{6}(k_1 + 4k_2 + k_3) + O(h^4) \quad (3)$$

Third order RK [7]

Without a doubt, the most frequently used is the conventional fourth order Runge-Kutta formula, which has a certain sleekness of orderliness about it:

$$k_1 = hf(x_n, y_n)$$
$$k_2 = hf\left(x_n + \frac{h}{2}, y_n + \frac{k_1}{2}\right)$$
$$k_3 = hf\left(x_n + \frac{h}{2}, y_n + \frac{k_2}{2}\right)$$
$$k_4 = hf(x_n + h, y_n + k_3)$$
$$y_{n+1} = y_n + \frac{k_1}{6} + \frac{k_2}{3} + \frac{k_3}{3} + \frac{k_4}{6} + O(h^5) \quad (4)$$

Fourth order RK [7]

The fourth order RK method requires four evaluations of the right hand side per step h. By so doing, it will be superior to the second order RK or midpoint method (Eq. 2); at least twice as large a step is achievable with (Eq. 4) for the same accuracy.

The assertion that fourth order RK is generally superior to second order is true, but one should know that it as an assertion about the contemporary practice of science rather than as an assertion about strict mathematics [8].

The above equations of RK are for calculating the first order ODE; below is the fourth order RK for second order ODE:

$$y' = f_1(x, y, z) \text{ where } z = y' \text{ and } z' = y''$$
$$z' = f_2(x, y, z)$$
$$y(x_0) = y_0, \ z(x_0) = z_0 \text{ are both initial conditions}$$

$$k_1 = hf_1(x_n, y_n, z_n)$$
$$l_1 = hf_2(x_n, y_n, z_n)$$
$$k_2 = hf_1\left(x_n + \frac{h}{2}, y_n + \frac{k_1}{2}, z_n + \frac{l_1}{2}\right)$$
$$l_2 = hf_2\left(x_n + \frac{h}{2}, y_n + \frac{k_1}{2}, z_n + \frac{l_1}{2}\right)$$

$$k_3 = hf_1\left(x_n + \frac{h}{2}, y_n + \frac{k_2}{2}, z_n + \frac{l_2}{2}\right)$$
$$l_3 = hf_2\left(x_n + \frac{h}{2}, y_n + \frac{k_2}{2}, z_n + \frac{l_2}{2}\right)$$
$$k_4 = hf_1(x_n + h, y_n + k_3, z_n + l_3)$$
$$l_4 = hf_2(x_n + h, y_n + k_3, z_n + l_3)$$
$$y_{n+1} = y_n + \frac{k_1}{6} + \frac{k_2}{3} + \frac{k_3}{3} + \frac{k_4}{6} + O(h^5)$$

$$z_{n+1} = z_n + \frac{l_1}{6} + \frac{l_2}{3} + \frac{l_3}{3} + \frac{l_4}{6} + O(h^5) \quad (5)$$

RK4 for second order ODE [3]

Runge-Kutta method is said to be unique based on the following properties such as:

- It is a one-step method which means that to find $y^{m+1}$, one would require information present at the preceding (initial) points: $x^m$ and $y^m$.
- It agrees with the Taylor's series through terms in $h^p$, where p differs for the various methods and called the order of the method.
- It does not require the evaluation of any derivatives but only requires the function itself.

Owing to the fact that RK is a method with different stages, it is rather called nth stage RK method. Where n=1, 2, 3, 4…..But for the purpose of this work, we shall limit ourselves to the fourth order RK method.

Angelos (2013) formulated using Artificial Neural Networks in the construction of Runge-Kutta methods by generating the optimal coefficients of a numerical method. The network was designed to produce a finite difference algorithm that solves a specific system of ordinary differential equations numerically. This majorly concerns an explicit two-stage Runge-Kutta method for the solution of the two-body problem numerically.

Following the implementation of the network, the latter is trained to obtain the optimal values for the coefficients of the Runge-Kutta method. The comparison of the new method to others that are well known in the literature proves its efficiency and demonstrates the capability of the network to provide efficient algorithms for specific problems.

Jorick (2005) presented an interesting form of flow problem, one that involves multiple fluids of flows; more particularly, the two-fluid flows where there exists two non-mixing fluids separated by a sharp fluid interface, occurring in many applications majorly in both engineering and physics. Although, the use of experimental and analytical results have provided us a solid foundation for two-fluid dynamics, research on two-fluid shows that areas involved with solving numerically the flow equations are otherwise known as Computational Fluid Dynamics (CFD).

However, we suppose that despite the disparity in the simulation of single-fluid flows, finding two-fluid flows that possesses accuracy and efficiency in the simulation method has proved abortive. This is due to the fact that when handling the interface between the two-fluids, difficulties often arise and also the numerical methods that manages these interface problems sometimes lack the accuracy. Therefore, the thesis presides the improvement of an extremely accurate numerical solver for the simulation of compressible, unsteady two-fluid flows as portrayed by the two-dimensional Euler equations of gas dynamics. The two-fluid flow solver considered was centered on the Level-Set (LS) method.

The uniqueness of the solver developed is the application of an extremely accurate Runge-Kutta discontinuous Galerkin (RKDG) method for the temporal and spatial discretization of the governing equations. The probability of gaining very high orders of accuracy and the somewhat easy implementation of mesh and order refinement techniques makes the RKDG method an appealing method for solving fluid flow problems. The RKDG method also relates the accuracy of the former with the efficiency and easy implementation of the latter of a two-fluid flow solver equally centered on the LS method and by itself results in an interesting numerical solver for two-fluid flows.

The development of high performance Runge-Kutta (RK) numerical methods applied to unravel the Schrodinger equation for Hydrogen and Positronium Atoms generated using numerical results which conforms to the analytical calculations of the hydrogen atom in modern physics and quantum mechanics based on the ground state modes of wave functions for both hydrogen and positronium. Numerical RK method to solve differential equations in physics is very efficient and accurate useful for solving physics problems and could also be used in the analysis of quantum systems with different potentials [5]

### 3. System Design

Several techniques are used to model systems such as the Unified Modeling Language (UML) which is a language for expressing object oriented design models. The UML diagram used for this research is the Use Case diagram.

3.1 Use Case Diagram

This is a diagram that shows the actors and use cases, together with the various relations between them. The use case for this system is represented in figure 3.1 below.

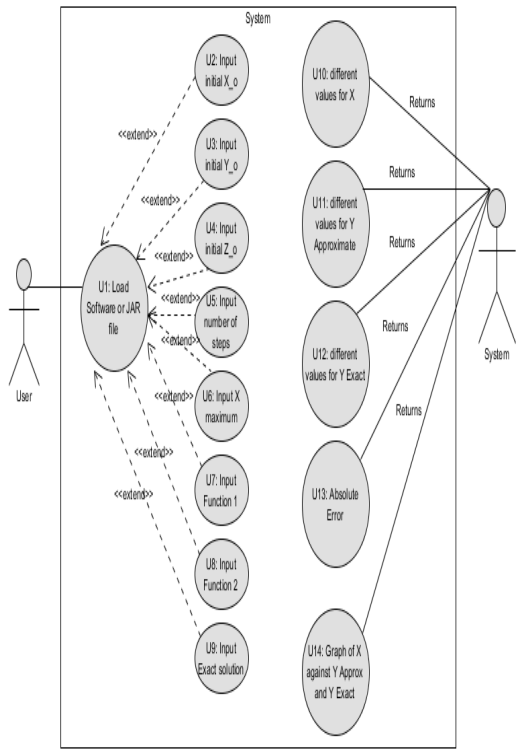

**Figure 1 Use Case Diagram for inputs into RK4**

3.2 Grammar Rules for Accepting Inputs

1. S ←⎯ S "+" or "– " E1
2. S ←⎯ E1
3. E1 ←⎯ E1 "*" or "/" E2
4. E1 ←⎯ E2
5. E2 ←⎯ E2 "^" E3
6. E2 ←⎯ E3
7. E3 ←⎯ "Sin" or "cos" or "tan" or "log" or "- " T
8. E3 ←⎯ T
9. T ←⎯ i
10. T ←⎯ (S)

Removing left recursion…….

1. S ←⎯ E1 Rs
2. Rs ←⎯ "+" or "– " E1 Rs | $E$
3. E1 ←⎯ E2 Re1
4. Re1 ←⎯ "*" or "/" E2 Re1 | $E$
5. E2 ←⎯ E3 Re2
6. Re2 ←⎯ "^" E3 Re2 | $E$
7. E3 ←⎯ "Sin" or "cos" or "tan" or "log" or "- " T
8. E3 ←⎯ T
9. T ←⎯ i
10. T ←⎯ (S)

To construct a parser based on this grammar, we use the recursive descent parser which is a kind of top-down parser. Some types of grammars especially the context-free grammars, allows a recursive descent parser to choose which production to use by considering only the next k tokens of input LL(K)- this grammar rules out all ambiguity present in the grammar as well as all grammars that are left recursive like the grammar above.

Based on this fact, a parser that will generate an error message if its input cannot be found in the language of the above grammar is constructed and also an Abstract Syntax Tree (AST) is constructed in order to correctly reflect the structure of the sequence of input.

## 4. Implementation Techniques

4.1 JAVA Programming Language

Java programming language was used to develop the front end of the Runge-Kutta fourth order system via the Java Development kit (JDK), which is the complier for Java programming language that converts Java source codes into executable Java Class files or libraries. Version 1.8.0 was used in the implementation

4.2 Steps in Implementing Dynamic Runge-Kutta Fourth Order Algorithm

1. Step 1: On the home page of Runge-Kutta fourth order, select "compute ODE".
2. Step 2: Enter the values for the corresponding fields.
3. Step 3: Select either first order or second order.
4. Step 4: Select 'Run'.
5. Step 5: A dialog box appears showing the chart and table.
6. Step 6: You can then select 'Exit' to go back to the main window

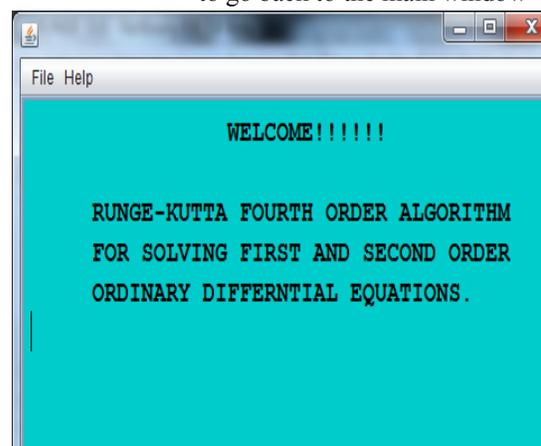

**Figure 2 Welcome screen for RK4 first and second order ODE**

**Figure 3 Data Entry Form**

After all the fields have been filled, one can choose to either select the "Run" button or "Exit". Note that the title of the field having "*" is not a compulsory field and can be ignored in the computation of first order ODE but is highly needed in the computation of second order ODE. Also, "#" indicates that when entering values into those fields, you terminate them with a semi-colon.

**Figure 4 Data entry to compute RK4 for (x-y)/2**

The above example computes the ODE (x-y)/2 which is a function of x and y with the initial value of x to be 0 and y to be 1; represented mathematically as: $y(0) = 1$. The number of steps to be determined is 24 and the maximum (end point) of x is 3. The solution to the ODE was analytically computed using the method of undermined co-efficient which reduces the solution to be a function of only x. The button 'first order' was then selected to run.

X represents the list of x values, Y_Approximate represents the list of y approximated values, Y_Exact represents the list of y exact (True solution) values and Absolute Error represents the difference between Y_Exact and Y_Approximate. The fixed step size denoted 'h' which has to do with how x increases is 0.125; is calculated by dividing the number of steps = 24 by the maximum value of x = 3.

## 5. Results

### 5.1 Screen Shots of Runge-Kutta Fourth Order Test Cases

#### 5.1.1 Example on First Order ODE

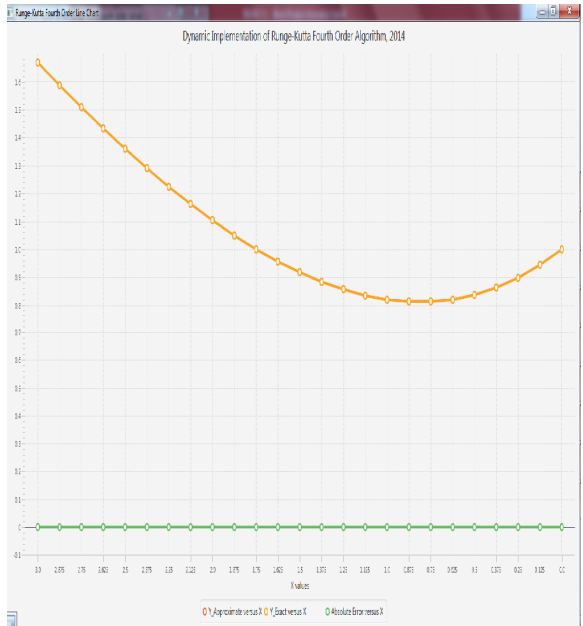

Figure 5 Graph of (x-y)/2function

In the graph above, observe that the horizontal values are reversed backwards. The colours red denotes Y Approximate, yellow denotes Y Exact and green denotes Absolute Error. This graph was plotted based on the tabulated values. We have plotted in this graph Y_Approximate against X, Y_Exact against X and Absolute Error against X. There is little or no difference at all between Y_Approximate values and Y_Exact values. As a result of this inference, we have in the graph above a single curve representing both Y_Approximate and Y_Exact as their values increases through X values and a straight line representing the Error. Also, notice that the absolute error is very small (minute) for all the values of x.

### 5.1.2 Example on Second Order ODE

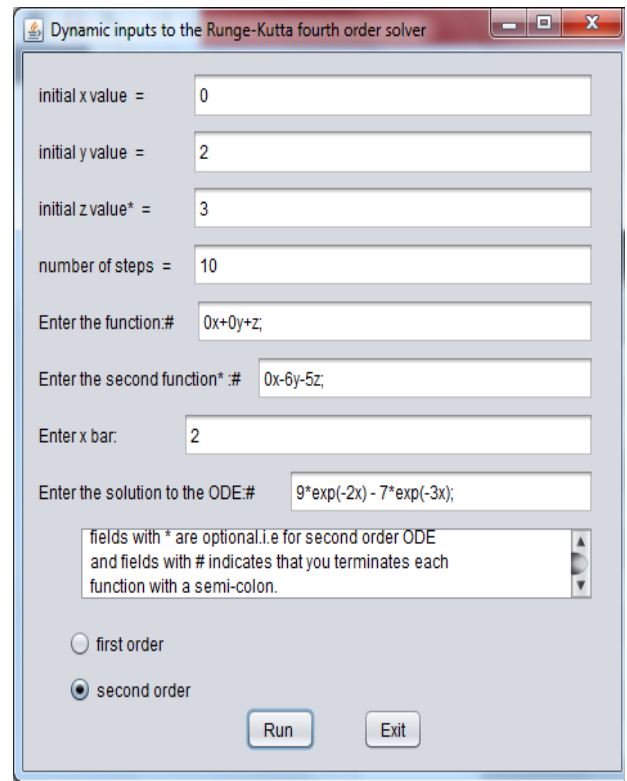

Figure 6 Data entry to compute RK4 for $y'' + 5\dot{y} + 6y = 0$ and $y' = z$

The above example computes the ODE $y'' + 5y' + 6y = 0$ which is a function of y and z with the initial value of x to be 0, y to be 2 and z to be 3; represented mathematically as:
$y(0) = 2, \ z(0) = 3$. For second order, the equations are divided into two:
$y' = z$ and $y'' = -6y - 5z$
The number of steps to be determined is 10 and the maximum (end point) of t is 2. The solution to the ODE was analytically computed using the method of undermined co-efficient which reduces the solution to be a function of only x. The button 'second order' was then selected to run.

X represents the list of x values, Y_Approximate represents the list of y approximated values, Y_Exact represents the list of y exact (True solution) values and Absolute Error represents the difference between Y_Exact and Y_Approximate.

The fixed step size denoted 'h' which has to do with how x increases is also 0.2; is calculated by dividing the number of steps = 10 by the maximum value of x = 2.

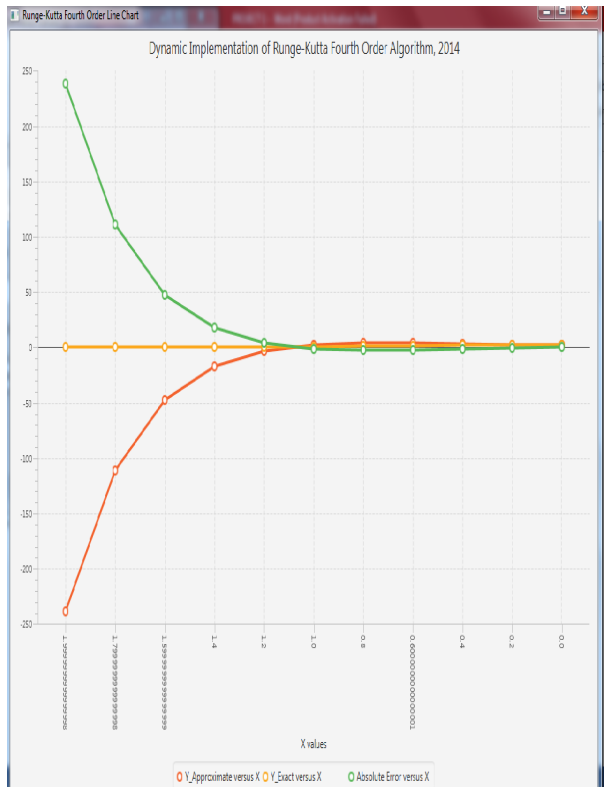

Figure 7 Graph of y′′ + 5 ẏ + 6 y= 0 function

We have plotted in this graph Y_Approximate against X, Y_Exact against X and Absolute Error against X. From the table, there is a difference between Y_Approximate values and Y_Exact values. As a result of this inference, we have in the graph above two different curves representing Y_Approximate and Absolute Error. The straight line in the graph represents Y_Exact. Also, observe that while the Absolute Error increases in the positive y-axis direction, Y_approximate increases in the negative y-axis direction.

## 6. Conclusion

This study shows that Runge-Kutta's fourth order algorithm can dynamically evaluate the inputs through the use of compiler techniques and can generate a solution for both first and second order ODE. This implementation is limited by the use of one programming language only. The Java program does not produce executable files, instead it compiles the program into bytecodes leaving it in a .jar format so the program has to be run from the command prompt window or console. Future research work may examine other algorithms as well as the introduction of Compilation techniques to be used in other areas of study particularly in the field of Numerical computations.

**Anidu Adesola.O.** holds a Masters degree in Computer Science and is currently on her Ph.D programme.  She is an assistant lecturer in the Department of Computer Science, Redeemer's University Ede. Her research interests include biometrics, mathematical computing, e-learning and artificial intelligence.

**Arekete Samson A.** holds a Ph.D degree in Computer Science. He has over 10 years experience in lecturing and is a senior lecturer in the Department of Computer Science, Redeemer's

University, Ede. His research areas are agents in mobile technology, intelligent agents technology, artificial intelligence and context-aaaware computing.

**Adedayo Ayomide O.** is a student in the Department of Computer Science, Redeemer's University. Her research interests are mathematical computing, compiler construction

**Adekoya Adekunle O.** holds a Masters degree in Computer Science and currently on his Ph.D programme. He is a senior programmer in Redeemer's University with over 10 years experience in programming.